\documentclass[10pt,final,journal,letterpaper,twoside,twocolumn]{IEEEtran}
\usepackage[pdftex]{graphicx}
\usepackage{epstopdf}
\DeclareGraphicsExtensions{.pdf,.jpeg,.png,.eps}
\usepackage{cite}
\usepackage{amssymb}
\usepackage{amsmath}
\usepackage{graphics}
\usepackage{graphicx}
\usepackage{float}
\usepackage[T1]{fontenc}
\usepackage[utf8]{inputenc}
\usepackage{authblk}
\usepackage{authblk}
\usepackage{tabularx}
\usepackage{balance}
\usepackage{ amssymb }
\usepackage{algorithm2e}
\usepackage{algorithmic}
\usepackage{mathtools}
\usepackage{gensymb}
\usepackage{subfig}
\usepackage{multicol}
\usepackage{lipsum}
\usepackage{chngpage}
\usepackage{calc}
\usepackage{ragged2e}
\usepackage{tabularx}
\usepackage{color}
\begin{document}
\title{Optical Non-Orthogonal Multiple Access \\ for Visible Light Communication}
\markboth{Submitted to the IEEE Wireless Communications Magazine}{Marshoud \MakeLowercase{\textit{et al.}}: Optical Non-Orthogonal Multiple Access for Visible Light Communication}

\author{
Hanaa~Marshoud,~Sami~Muhaidat,~Paschalis~C.~Sofotasios,~Sajjad~Hussain,\\~Muhammad~Ali~Imran,~and~Bayan~S.~Sharif
\thanks{H.~Marshoud  and B. S. Sharif  are with the Department of Electrical and Computer Engineering, Khalifa University, PO Box 127788, Abu Dhabi,  UAE (email:  $\{\rm hanaa.marshoud; bayan.sharif \}@kustar.ac.ae$).}

\thanks{S.~Muhaidat is with the Department of Electrical and Computer Engineering, Khalifa University, PO Box 127788, Abu Dhabi, UAE, and with the Institute for Communication Systems, University of Surrey, GU2 7XH, Guildford, UK (email: muhaidat@ieee.org).}

\thanks{P. C. Sofotasios is with the Department of Electrical and Computer Engineering, Khalifa
University, PO Box 127788, Abu Dhabi, UAE, and  with the Department of Electronics and Communications Engineering, Tampere University of Technology, 33101 Tampere, Finland (e-mail: p.sofotasios@ieee.org).}

\thanks{S.~Hussain and M. A. Imran are with the School of Engineering, University of Glasgow,  Glasgow G12 8QQ,  UK  (e-mail: $\{\rm sajjad.hussain;muhammad.imran\}@glasgow.ac.uk$).}

}
\maketitle
\begin{abstract}
The proliferation of mobile Internet and connected  devices, offering a variety of services at different levels of performance, represents a major challenge for the fifth generation  wireless networks  and beyond. This requires a paradigm shift towards the development of key enabling techniques for the next generation wireless networks.  In this respect, visible light communication (VLC) has recently emerged as a new communication paradigm that is capable of providing  ubiquitous connectivity  by  complementing  radio frequency communications. One of the main challenges of VLC systems, however,  is the low modulation bandwidth of the light-emitting-diodes,  which is in the megahertz range.   This article presents  a promising technology, referred to  as "optical- non-orthogonal multiple access (O-NOMA)",  which is envisioned to address  the key challenges  in the next generation of wireless networks.  We provide a detailed overview and analysis of the state-of-the-art integration of O-NOMA in VLC networks.  Furthermore,  we provide insights on the potential opportunities and challenges as well as some open research problems that are envisioned  to pave the way for the future design and implementation of O-NOMA  in VLC systems.
\end{abstract}
\section*{Introduction}
The explosive growth of connected devices, due to the emergence of Internet of Things (IoT) and the growing number of broadband mobile subscribers, which is expected to be around   8.6 billion subscribers by 2020 [1],  will lead to an unprecedented growth in traffic demand. In this respect, the next generation wireless networks are envisioned to meet this growth and offer a projected data rate of 20 Gbps, posing new technical challenges to address  the requirement for low latency and high spectrum efficiency. The ongoing research efforts have mainly focused on two main directions: 1) enhancing the spectral efficiency of the available radio frequency (RF) spectrum by adopting advanced modulation schemes, new multiple access techniques and efficient bandwidth reuse; and 2) exploring the potentials  of the unlicensed spectrum, i.e.,  Infrared (300 GHz - 430 THz) and visible light spectrum (430 THz - 790 THz).   In this context, visible light communication (VLC)  has emerged as a promising  small cell technology that can be connected to  the existing super fast fiber networks and constitutes  an integral part of the future ubiquitous fifth generation (5G)  communication systems.

A VLC system uses off-the-shelf standard light-emitting diodes (LEDs) to enable high speed data transmission and, simultaneously, to provide indoor/outdoor illumination.  It can be realized by modulating and demodulating  the light intensity of the LEDs in a process  known as  intensity modulation and direct detection (IM/DD).  The most attractive features of VLC are the  inherent communication security, the  high degrees of spatial reuse,  and its invulnerability  to RF interference, which   renders   it safe for operation in environments  with high electromagnetic interference (EMI), such as  hospitals and industrial plants. Furthermore, visible light cannot penetrate through most objects and walls, making it well-suited for small cell design and capable of  providing high quality services without inter-cell interference. This constitutes  VLC  a viable attractive technology and an effective  complement
to current RF communications.

Nonetheless, in spite of its great advantages, VLC  systems have certain shortcomings  which need to be fully addressed in order to exploit the  full potentials of this emerging  technology.  A key limitation of  VLC systems is that the achievable data rates are restricted to the limited modulation bandwidth of the LEDs, which spans few MHz. Therefore, in order to realize the envisioned VLC systems with full potentials, there has been  increased interest in the efficient development  of advanced optical modulation techniques, optical multiple-input-multiple-output (MIMO) and multiple access schemes.

As a technology enabler for 5G networks and beyond, VLC is envisioned to provide  ubiquitous  connectivity and high spectral efficiency. To this end, various  optical (orthogonal and non-orthogonal) multiple access schemes have been proposed  in the open literature, addressing these challenges.  In orthogonal multiple access (OMA),  different users are allocated to orthogonal resources in either the  frequency or time domain. For example, orthogonal frequency division multiple access (OFDMA) assigns different frequency sub-carriers  to different users, whereas time division multiple access (TDMA) allows users to share the same frequency by accessing the network  in a rapid succession  during their assigned time slots.

On the contrary, non-orthogonal multiple access (NOMA) allows multiple users to simultaneously utilize the entire available frequency and time resources,  leading to superior enhancements in   spectral efficiency compared to OMA schemes. NOMA can be realized via  two different approaches, namely, power-domain (PD) NOMA and code-domain (CD) NOMA. In PD NOMA,  users are multiplexed in the power domain by assigning distinct  power levels to  different users, while CD  NOMA utilizes user-specific spreading sequences in order to multiplex the users in the code domain similarly to   the well known code division multiple access (CDMA) technique.

It is worth noting that   PD NOMA  has been  shown to be particularly suitable  for VLC downlink (DL) systems for several  reasons \cite{NOMA_VLC}. First,
PD NOMA is typically used to multiplex  a few number of users, which is  in-line with the requirements of VLC systems that employ LEDs as small cells to serve a small number of users.
Second, power allocation in PD NOMA is exclusively dependent on  the  channel state information (CSI) available at the transmitter, and channel estimation in VLC is considerably less error-prone, compared to RF, due to the time-invariant channel gain that remains constant until the location of the receiving terminal changes.
Third, PD NOMA performs better in high signal-to-noise ratio (SNR) scenarios, which is  the case in VLC systems due to the strong line-of-sight (LOS) channel gain and the short separation between the communication front ends.
Finally, the performance of PD NOMA systems can be improved  by enhancing  the channel gain differences between users, which can be achieved  by tuning the transmission angles of the LEDs and the field-of-views (FOVs) of the PDs, offering two degrees of freedom to optimize the system performance. \emph{Henceforth, we refer to PD NOMA in VLC, which is the focus of this article,  as optical (O)-NOMA}.

\section*{Basic Concepts of O-NOMA}
\label{sec:concepts}
\begin{figure}[h]
\includegraphics[width=3.5in,height=3in]{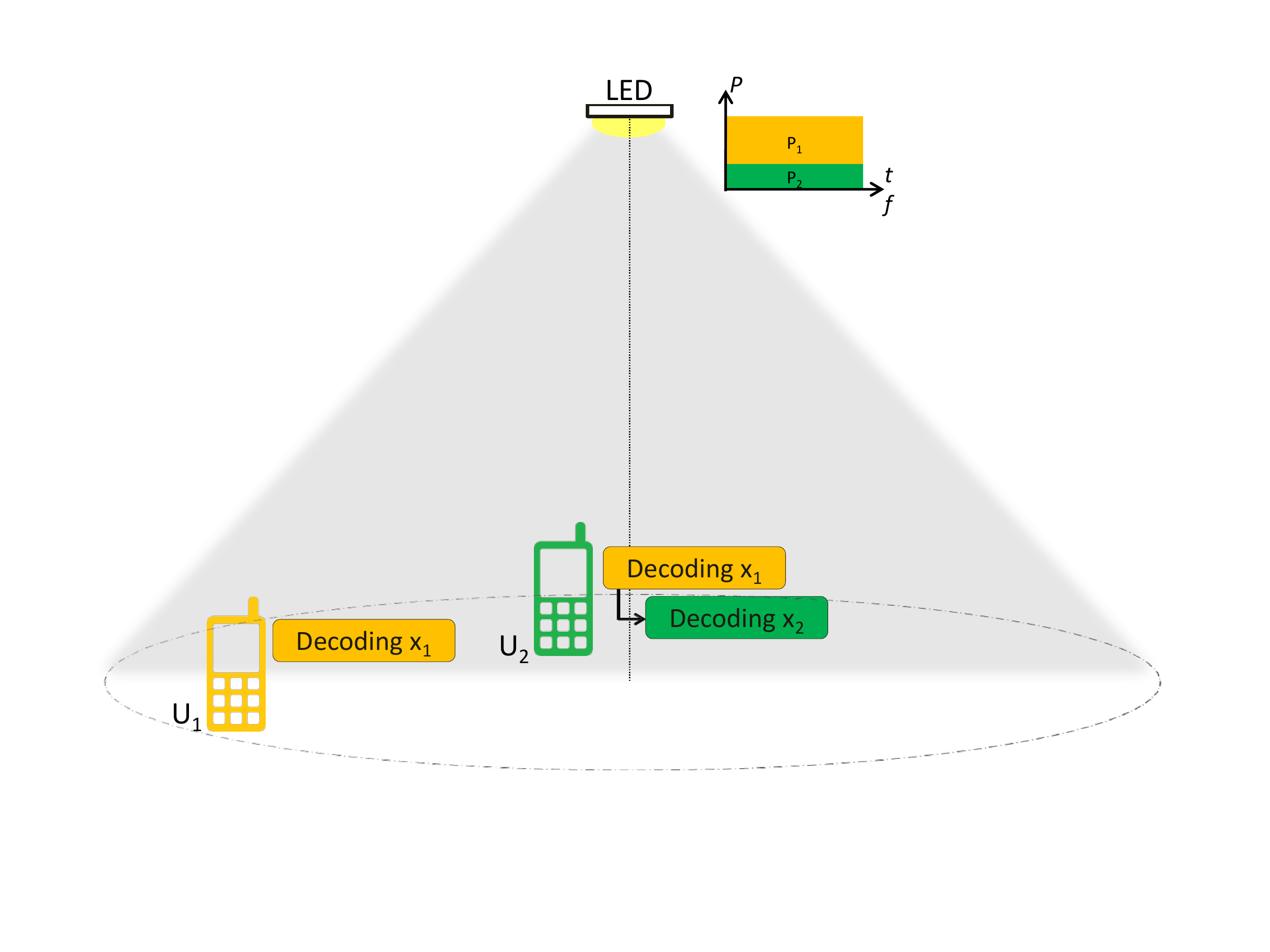}
\caption{Downlink NOMA VLC with two users.}
\label{fig:NOMA}
\end{figure}

O-NOMA allows different users to share the same time and frequency resources in the power domain,  leading to enhanced  capacity gains. Fig. \ref{fig:NOMA} illustrates the main components of  a two user O-NOMA   with a successive interference cancellation (SIC)  receiver.  In O-NOMA, power-domain multiplexing is  performed by  assigning different power values to the signals of the different users and transmitting them simultaneously. This process is referred to as superposition coding (SC) while multi-user detection is realized   using  SIC at the receiving terminals. In the following subsections,  we provide an overview of the  basic concepts of  O-NOMA.
\subsection*{Superposition Coding}
The concept of SC was first introduced  in \cite{SC},  where a transmitter can  simultaneously convey different information signals to  several receivers in a downlink broadcast channel. In O-NOMA, SC is realized by allocating high power values to  users with unfavorable channel conditions and vice versa.  In Fig. \ref{fig:NOMA}, the LED transmits the unipolar real signals $x_1$ and $x_2$ to $U_1$ and $U_2$, respectively. Since $U_2$ is closer to the transmitting LED and  thus, has a higher channel gain, the access point  assigns a lower power level to $x_2$. The two signals are then superimposed and transmitted simultaneously as $s=P_1 x_1 + P_2 x_2$, where $P_1 > P_2$ and the sum of the assigned power values is equal to the total LED transmitting power. The same principle applies for higher number of users, where the allocated power values are determined based on the channel gains of the different users.
\subsection*{Successive Interference Cancellation (SIC)}
Since the dominant component in the combined received signal in Fig. \ref{fig:NOMA}  is $P_1 x_1$, $U_1$ can directly decode its signal  by treating the interference from the other user's signal as noise.  However, $U_2$ must  decode $x_1$ first, and then subtract it from the combined signal in order to isolate $x_2$ from the residue. This process is called SIC where   the involved    users are ordered according to their respective signal strengths,  so that each receiving terminal decodes the   strongest signal first, subtracts it, and repeats the process until it decodes its desired signal.

\label{sec:results}
\begin{figure*}[   ]
\normalsize
\subfloat[]{\label{ref_label1}\includegraphics[width=2.5in,height=2.4in]{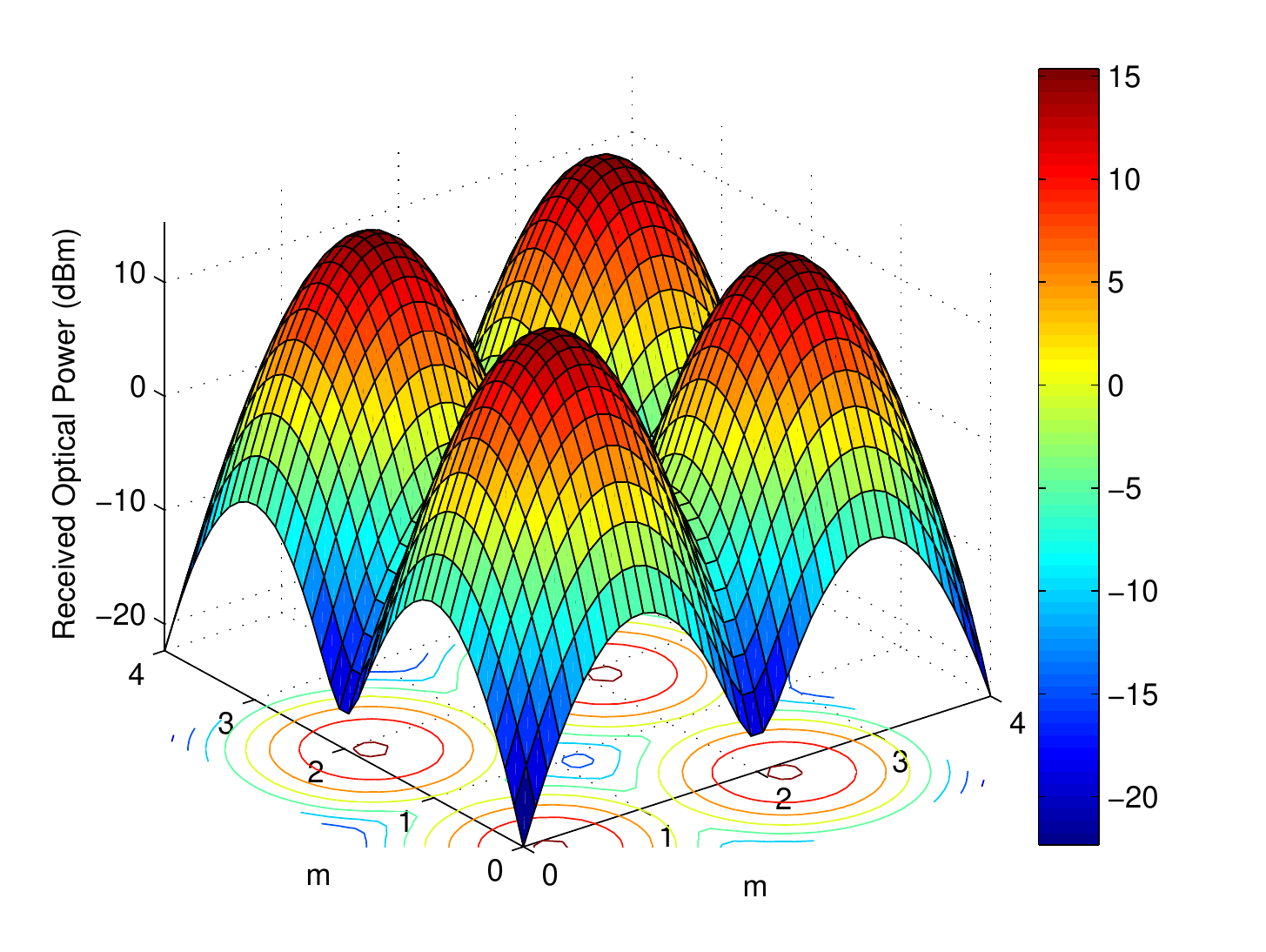}}
  \subfloat[]{\label{ref_label2}\includegraphics[width=2.5in,height=2.4in]{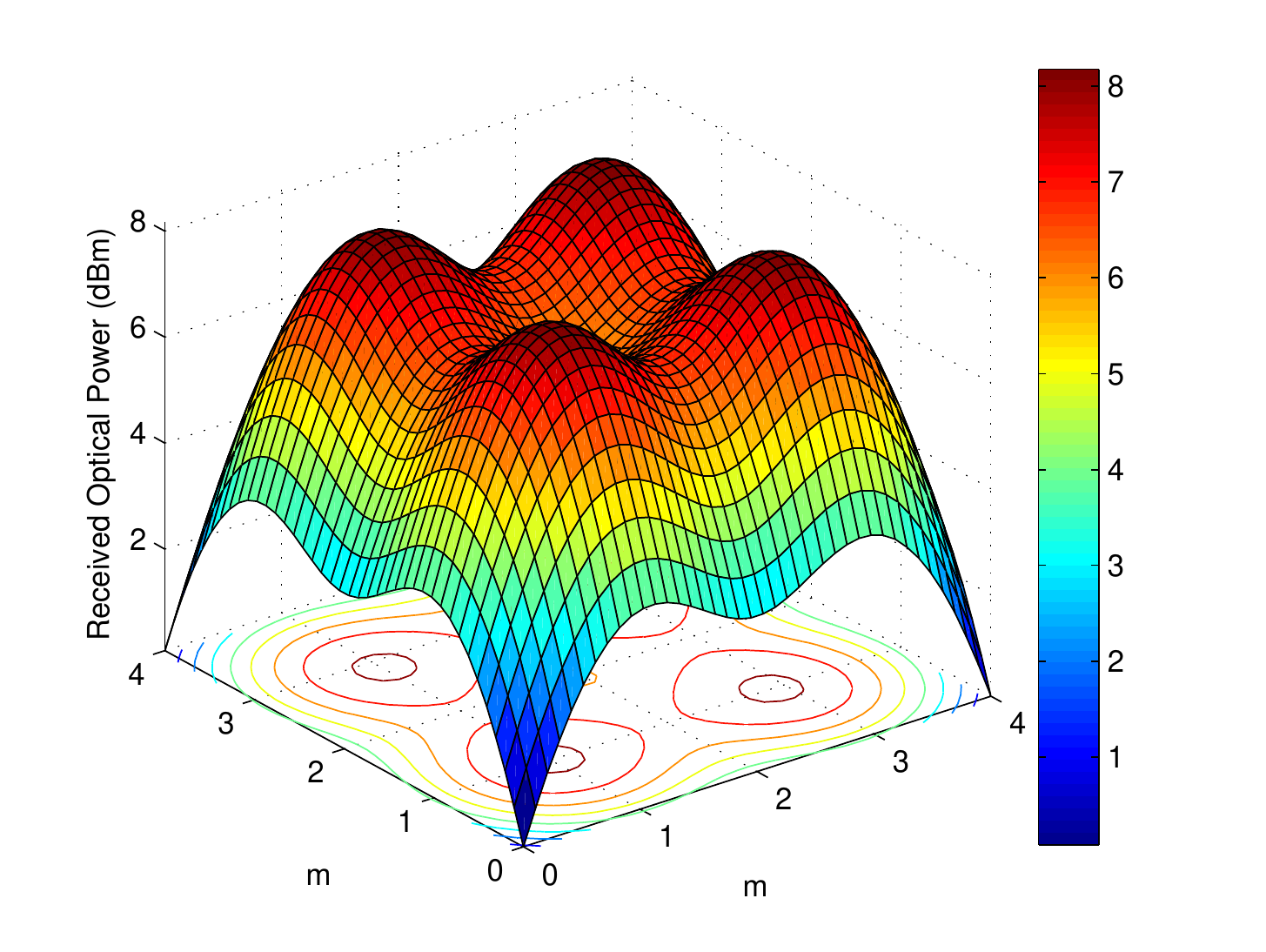}}
  \subfloat[]{\label{ref_label2}\includegraphics[width=2.5in,height=2.4in]{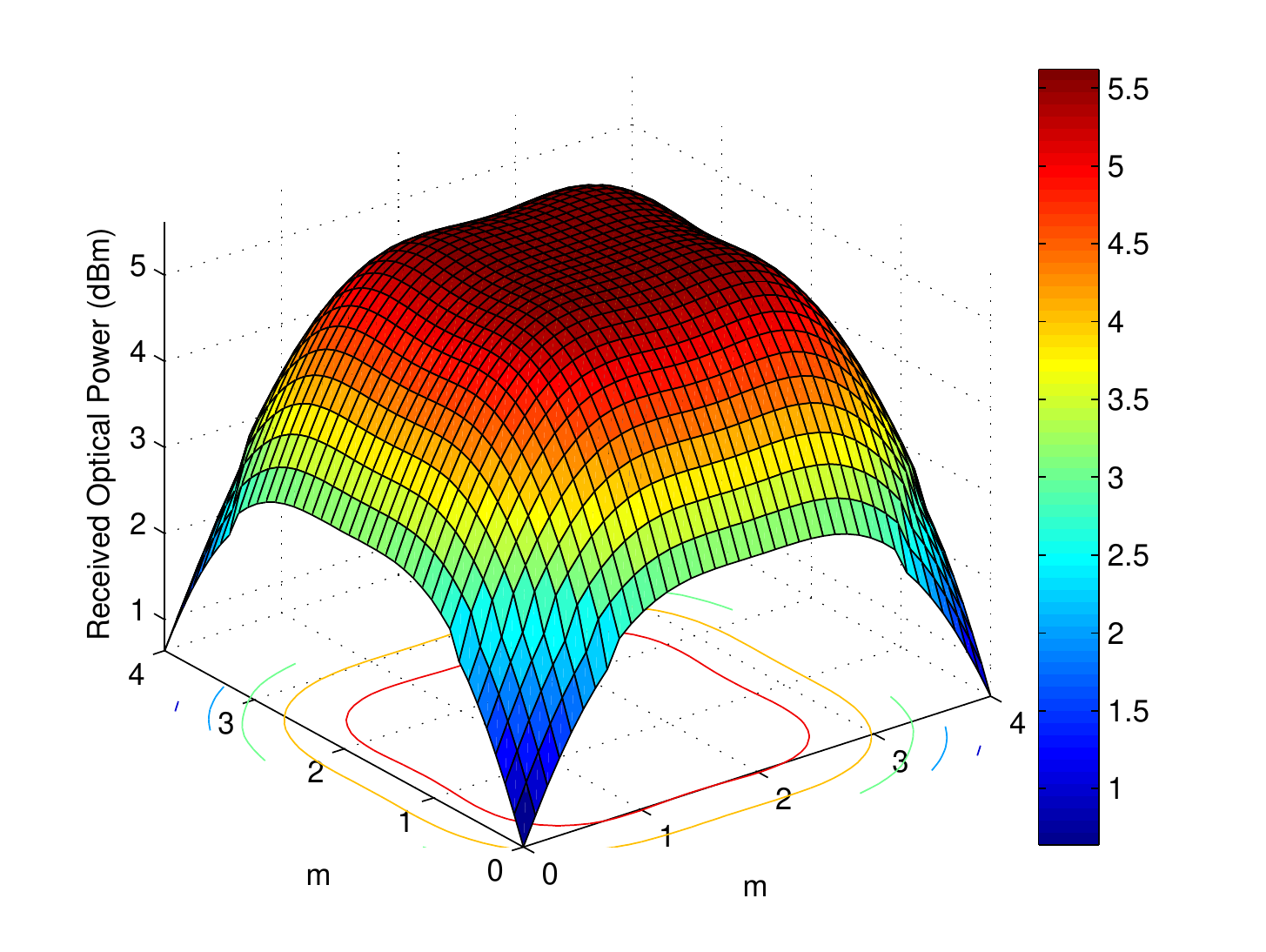}}
  \caption{\label{fig:h}Optical Power Distribution  for four transmitting LEDs with  (a) $\varphi_{1/2}=10^\circ$, (b) $\varphi_{1/2}=25^\circ$ and (c) $\varphi_{1/2}=45^\circ$.}
\noindent\makebox[\linewidth]{\rule{520pt}{.4pt}}
\end{figure*}

\subsection*{Power Allocation in O-NOMA}

O-NOMA allocates more power to users with worse channel conditions, while less power is allocated to those with better channel conditions.   A key element in this approach   is to allocate appropriate power levels for the different users in order to facilitate SIC and to achieve  better trade-off between throughput and fairness.  The simplest power allocation strategy is the so-called fixed power allocation (FPA) \cite{NOMA_VLC2}, in which  users are sorted in  an ascending order according to channel gain values. The power assigned to the $i^{th}$ sorted user is calculated as $P_i = \alpha P_{i-1}$, where $0<\alpha <1$. It is noted that FPA  does not require the exact channel gain values of the users, but rather their respective ordering which mainly depends on the users' distances from the transmitting LED.

In  \cite{NOMA_VLC}, a gain ratio power allocation (GRPA) strategy was proposed, where the power assigned to the $i^{th}$ sorted user is given by  $P_i = {({h_1}/{h_i})}^i P_{i-1}$. Since GRPA takes into account the actual channel path gains of all users, it was shown to provide better performance compared to FPA.  The performance of O-NOMA systems can be further enhanced by optimizing the power values according to   the users' specific channel conditions.  Exhaustive search was conducted in \cite{NOMA_VLC1} in order to determine  the optimum set of power coefficients that maximized the coverage probability of O-NOMA  systems. Similarly, an optimal power allocation algorithm was proposed in \cite{NOMA_VLC3}, where the allocated power coefficients were optimized in order to   maximize the sum throughput of the multi-user VLC system. The proposed algorithm was shown to outperform FPA and GRPA in terms of the achievable system sum rate. A similar approach was adopted in \cite{NOMA_VLC4} where the assigned power coefficients were dynamically optimized under quality of service (QoS) constraints to maximize sum rate or max-min  rate.

\section*{Multi-Cell  O-NOMA Networks}
\label{sec:multicell}
To support the explosive growth of mobile data, future wireless networks will continue to evolve, becoming smaller in size  and eventually  ultra dense to offload and localize traffic and increase the system spectral efficiency. In the next decade, it is expected  that ultra dense networks, with small cells, will cover most of indoor and outdoor spaces, providing data rate of 100 Mbps  to cell edge users \cite{nokia}. However, this will result in an increase in frequency reuse and will  introduce intolerable interference, limiting the spectral efficiency of the system. In this respect, VLC is envisioned  to play  a vital role in addressing these challenges, owing to  its unique characteristics.

A typical indoor environment  usually  comprises multiple adjacent  LEDs which form bordering or overlapping VLC cells. The  VLC cell size is determined by the transmitting angle of the LED and the vertical distance between the LEDs' plane and the plane of the receiving terminals. Fig. \ref{fig:h} shows the optical power distribution in a  $4\times4\times3$ m$^3$ room with four transmitting LEDs aligned  quadratically in a   $2\times2$ array separated by $1$ m and centered in the middle of the room, for different values of LED transmitting angle,  $\varphi$ . As shown in Fig. \ref{fig:h}, small transmitting angles form narrow beams with clear crisp borders,  whereas larger transmitting angles diffuse   the transmitted power  leading to seamless coverage area. The use of large transmitting angles is useful in broadcast transmissions, e.g. when all users are receiving the same information in  a museum or in  an exhibition. However, for scenarios with unicast transmissions, the overlapping of the VLC cells causes inter-cell interference, which raises the need for   sophisticated interference cancellation techniques.
\begin{figure}[h]
\includegraphics[width=3.5in,height=3in]{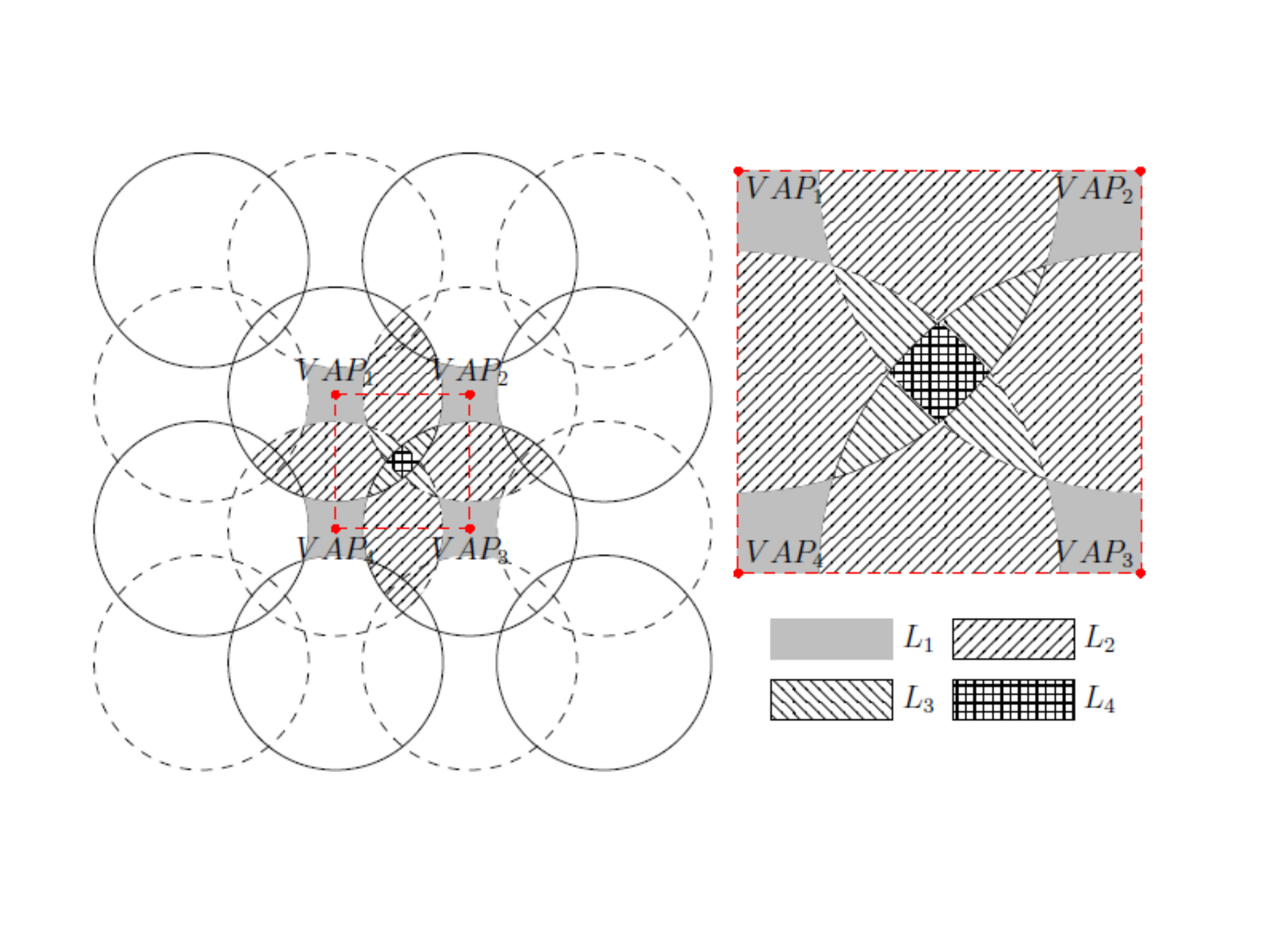}
\caption{Multi-cell NOMA VLC network \cite{NOMA_VLC4} }
\label{fig:cells}
\end{figure}
\subsection*{Inter-Cell Interference}

Location-based user grouping was adopted in \cite{NOMA_VLC4}  to mitigate the inter-cell interference in a multi-cell O-NOMA network. The coverage areas were classified into different types according to the degree of the cell overlapping,  as illustrated in Fig. \ref{fig:cells}, where VAP stands for VLC access point. Frequency reuse factor of two  was implemented to balance the trade-off between interference cancellation and spectral efficiency, where the bandwidth was divided among two groups of LEDs (represented by solid and dashed lines).  Also, user  grouping was  realized  as follows: users in area type $L_1$ received from a single LED without interference. Users in $L_2$ received from two overlapping LEDs; however, each LED used  a different part of the spectrum, so no interference existed. Moreover, users in area type  $L_3$ could receive from two different LEDs operating in the same bandwidth, so they   were  scheduled for load balancing. For users in area type $L_4$, a dedicated part of the bandwidth was reserved since these zones exhibited  interference from four different LEDs.

The problem of overlapping VLC cells was further investigated in \cite{NOMA_VLC5}, where a user located in the coverage area of two adjacent LEDs received two signals superimposed in the power domain with different power intensities, depending on the observed channel gains. The user then performed SIC to detect the individual signals from the composite constellation. It was shown that the system performance can be enhanced by applying phase pre-distortion to   the transmitted signals to achieve the optimal phase difference between the two channel gains.

A cell zooming approach was proposed in \cite{NOMA_VLC} to avoid cell overlapping, which was performed by adjusting  the transmitting angles of the LEDs so as  to control the respective cell sizes. In the proposed framework, each LED had two different transmission angle settings. A central control unit collected  the necessary information about users' locations to configure  the transmitting angles accordingly. It is noted here that  it is not always feasible to shrink the coverage areas of the LEDs as this may lead to grey holes, i.e., areas with no coverage. Moreover, the adjustment of the transmitting angles may affect the width and intensity of the light beams  leading to undesired  illumination  inconsistency across the indoor space.
\subsection*{Users Association and Handover}
In order to ensure seamless user experience in multi-cell NOMA configurations, a location-based  users association strategy was proposed in \cite{NOMA_VLC}. In particular, a user located in an overlapping area of two adjacent cells was instructed to remain connected to both LEDs, further enhancing the performance of  cell edge users. Moreover, the  FOVs of the receiving PDs were exploited  to reduce  the number of handovers. To this end, cell-edge mobile users used   wider FOV settings in order to remain  connected to their respective home cells for longer periods, avoiding unnecessary handovers. It is worth noting here that a large FOV setting lowers  the channel gains at the receiving terminal and,  thus,  the respective received SNR is reduced.
\section*{Hybrid OFDMA/TDMA-O-NOMA}
\label{sec:hybrid}
The capacity gain of O-NOMA is achieved by multiplexing different users in power domain while sharing same frequency and time resources. Due to interference constraints, it is impractical to multiplex a large group of users, which is consistent  with the nature of VLC systems. This  indeed stems from the fact that an LED is regarded as a small cell with a limited number of users existing within a small coverage area. Recent results have demonstrated   the multiplexing of up to $5$ users using a single LED transmitter. We point out   that for larger   numbers of users, particularly,   in a small  cell size, the channel conditions  may not differ
significantly  among users. In this case,  OMA can be a better choice and, therefore,  hybrid OMA and NOMA can coexist   to fulfill  a better  trade-off between capacity and reliability. To this end, users can be divided into different groups that are multiplexed via OMA technique such as OFDMA or TDMA, whereas the users within  each group are multiplexed in the power domain via O-NOMA.

It is evident that  the performance of such hybrid schemes is highly affected by the  user selection
strategy. The impact of users pairing in a hybrid multiple access  VLC downlink network was investigated in \cite{NOMA_haas}, where   users were divided into groups of two and a channel gain-based  pairing strategy was adopted. It was shown that the achieved system throughput can be maximized  by pairing the two users with the most distinctive channel conditions. It is then clear  that the choice of users pairing is not a straightforward problem as we cannot simply pair the users with the most dissimilar channels, leaving users with correlated channels to suffer interference-limited performance. Therefore, optimum user-pairing  in hybrid multiple access systems  requires  sophisticated algorithms to obtain the maximum benefits offered by O-NOMA.   Furthermore, although indoor VLC systems typically exhibit  low mobility velocity, any small change in the users' locations would change the corresponding  channel gain,  due to the short distance and the nature of the VLC channel. As a result, low complexity dynamic users-pairing algorithms  are required  for  practical system implementations.
\section*{Performance of O-NOMA}
\label{sec:performnce}

The latest research efforts on the  performance evaluation of O-NOMA  have mainly focused on the capacity gains of O-NOMA, compared to its OMA counterparts. However, the  capacity gain naturally comes at the expense of reduced link reliability. It is evident that splitting the power between users leads to a lower received SNR and, consequently, higher error probability. Moreover, the inherit interference implied  by power-domain superposition and the cancellation errors that may occur  during SIC lead to lower detection accuracy. In this section, we provide an overview of the performance measures  of O-NOMA  systems, and we develop useful  insights into  the inevitable  tradeoff between capacity and reliability.
\subsection*{System Capacity}
The work in \cite{NOMA_VLC1} and \cite{NOMA_haas} provided  analytical and numerical  evaluation of  the ergodic sum rate  for different number of users, where the individual data rates were  opportunistically assigned  in a best-effort manner based on the users' channel conditions. In the aforementioned study, O-NOMA showed superior performance in terms of the system sum rate compared to  OFDMA. Interestingly, it was shown that although the blockage of the LOS channel component degrades the system performance, O-NOMA transmission was still possible due to the existence of multipath reflections. The sum rate performance of a multi-cell O-NOMA  network  was investigated in \cite{NOMA_VLC4}. It was shown that employing a frequency reuse factor of two  can lead to significant improvement in the achievable   sum rate in each cell compared to the case without frequency reuse. Moreover, it was shown that CSI error may lead to degradation in the achieved sum rate of O-NOMA, as it affects the accuracy of the power allocation.  Nevertheless,  O-NOMA sum rate  remained   higher than the sum rate of OMA, even with imperfect CSI.

DC-biased optical OFDM (DCO-OFDM)  transmissions were considered in \cite{NOMA_VLC2} for both O-NOMA and OFDMA. It was shown that O-NOMA outperforms OFDMA in terms of the achievable sum rate under zero or low cancellation errors. However, for higher cancellation errors, which may result from inefficient power allocation,  the performance of O-NOMA suffered severe performance loss.
\subsection*{Link Reliability}
The coverage probability of an O-NOMA  system was investigated in \cite{NOMA_haas}, where it was defined as the probability that all users in the system achieve a reliable detection. For optimally chosen power coefficients,   the system coverage probability was nearly 100\%  for  low target data rate, and it decreased  with the increase in the target data rate. Compared to OFDMA, O-NOMA provided higher coverage probability for different target data rates. 

The  error  performance of O-NOMA  systems was analyzed in  \cite{NOMA_HANAA} for the case of perfect and imperfect CSI. The derived  bit-error-rate (BER) expressions explicitly  took  channel uncertainty, cancellation errors, and interference terms into account. It was shown that noisy CSI, as expected, leads to  a degradation of the system performance. This degradation is rather small compared to the one created by outdated CSI which may result from the mobility of the user between two consecutive  CSI updates. In fact, outdated CSI can cause detrimental performance loss if the ordering of the users’ channel gains change between the channel updates, leading to unfair power allocation.
\subsection*{Performance Trade-offs}
\begin{figure}[h]
\includegraphics[width=3.5in,height=3in]{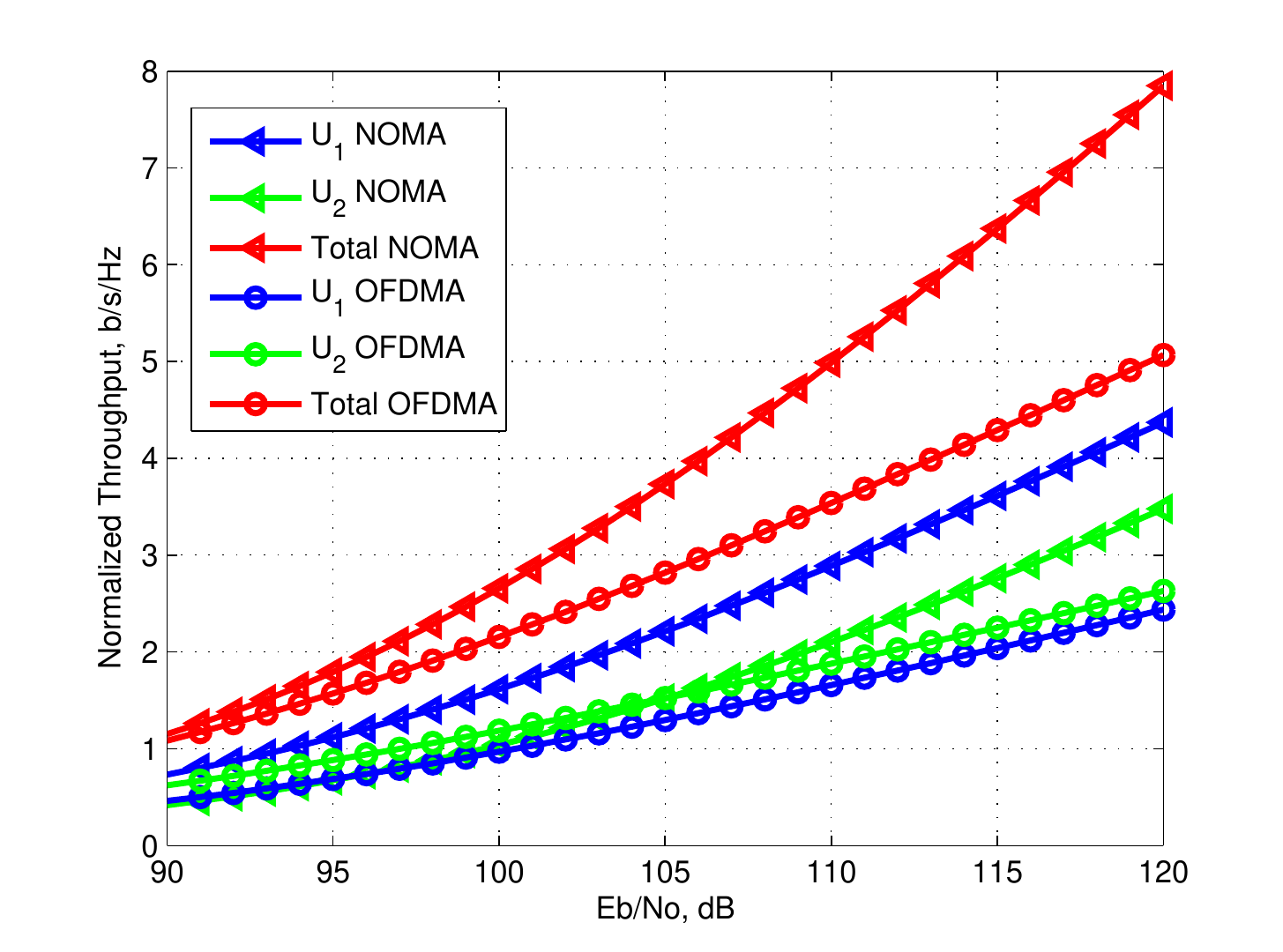}
\caption{Normalized System Throughput for Two Users}
\label{fig:rate}
\end{figure}

\begin{figure}[h]
\includegraphics[width=3.5in,height=3in]{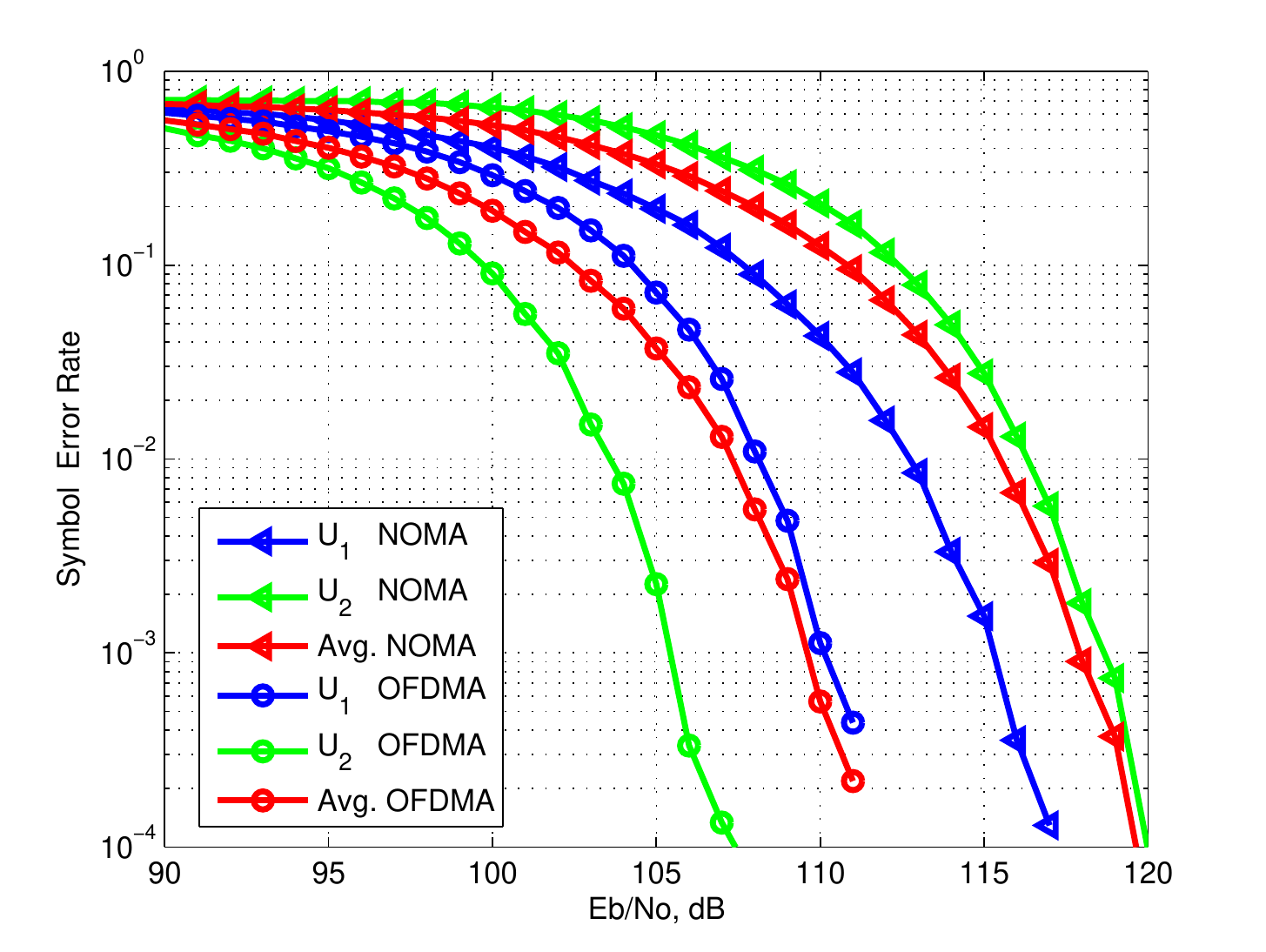}
\caption{Error Rate Performance for Two Users}
\label{fig:ser}
\end{figure}

In order to demonstrate the trade-off between capacity and reliability, we simulate the performance of  a two-user scenario for both O-NOMA and OFDMA.   Fig. \ref{fig:rate} and Fig. \ref{fig:ser} show the normalized system throughput and the error rate performance,  respectively. Both multiple access schemes are based on DCO-OFDM transmission and $U_1$ indicates the user with a lower  channel gain. As can be seen from Fig. \ref{fig:rate}, O-NOMA provides a significant improvement  in the overall system throughput  compared to OFDMA. This throughput gain is higher for $U_1$, since O-NOMA allocates higher power to  users with unfavorable channel conditions. However, the observed throughput enhancements provided by O-NOMA come at the expense of performance loss in  link reliability, as demonstrated by  Fig. \ref{fig:ser}. As  expected,  O-NOMA suffers from  higher  error rate, particularly,  for $U_2$, since it performs signal detection in  the presence of interference from the signal of $U_1$.

\section*{Future Research  Directions}
\label{sec:future}
So far, we have introduced the ongoing research efforts  on implementing O-NOMA. In the following, we identify some interesting
research opportunities and challenges related to the  practical integration  of NOMA in VLC systems.
\subsection*{MIMO-O-NOMA}
In order to ensure  sufficient illumination, indoor spaces are typically equipped with multiple LEDs. This feature has motivated the implementation of  MIMO  configurations in VLC systems \cite{MIMO_VLC_Haas}. MIMO can be used in conjunction with NOMA to provide capacity and reliability improvements.  For MIMO with spatial multiplexing (SMP), users can be divided into subgroups where each group  receives from a single LED. Within each group, users are superimposed in the power domain; this implies that two levels of interference cancellation are
needed:  the first level is to separate the  MIMO subchannels, which can be attained by means of transmit precoding \cite{marshoud};  the second is the interference cancellation  among the multiplexed users of the same LED, which can be performed using SIC. A careful design of the SMP-MIMO NOMA can provide significant capacity enhancement by increasing the number of users that are  simultaneously served.  On the contrary,  MIMO-based repetition coding (RC)   can be used to improve link reliability by providing diversity gains. However, the design of such system is not straightforward when O-NOMA is considered. This is because the power allocation should be performed across the aggregate signal depending on the users' channel gains from the different LEDs, which increases the involved complexity  specially when users' mobility is considered.

\subsection*{Impact of Channel Symmetry}
It has been pointed out earlier that the existence of dissimilar channel gains is the basis of successful detection in O-NOMA. However,  the VLC channel implies that the users' channel gains  could be particularly  close to one another, or even equal,  depending on the positions of the receiving terminals. Typical indoor VLC configurations exhibit strong symmetrical channel gains as the LEDs are located in the center of the room, which forms a real challenge in O-NOMA. One possible solution is the adjustment  of the transmitters' angles and the receivers' FOVs, as reported in \cite{NOMA_VLC}.  However, further research is needed to evaluate the performance of O-NOMA   in highly symmetrical setups, and to investigate how angles adjustment  can be dynamically performed when users' change their locations.

\subsection*{Nonlinear Distortion in O-NOMA}

Nonlinear distortions  constitute  limiting factors of the performance of  VLC systems, although  they have been overlooked in the majority of recent literature. The most common sources of nonlinearities  are the  circuits of the LEDs, LEDs, photodiodes, and digital-to-analog/analog-to-digital converters. Furthermore, VLC systems are  sensitive to nonlinearities in  electrical-to-optical  and optical-to-electrical conversion  \cite{nonlinear}.
The  impact of nonlinearities on the performance of O-NOMA  is not comprehensively understood yet,  since it has not been addressed  in the related open literature, which demands for  a thorough investigation. We note that such an investigation is imperative for  the actual realization  of O-NOMA and for determining the  actual performance limits in terms of both spectral efficiency and error rate performance.  Finally, the design of  pragmatic compensation strategies to mitigate  nonlinear distortions constitutes an important challenge 	 for future high speed VLC systems design, which needs to be addressed  before O-NOMA becomes an integral part of future wireless networks.

\subsection*{Other Challenges}
Future research and development of  O-NOMA in the context of VLC can be  directed towards  the    implementation of O-NOMA in Massive MIMO VLC \cite{massive-mimo1}, proper users pairing and power allocation techniques under feedback delay, peak-to-average-power ratio (PAPR) reduction in hybrid OMA-O-NOMA systems,    O-NOMA in asynchronous communications, and O-NOMA in uplink VLC.

\section*{Conclusions}
\label{sec:conc}
This article reviewed  the emerging concept of power-domain O-NOMA and its integration in VLC systems. A critical review of the state-of-the-art on the design and implementation of O-NOMA VLC allowed us to recognize the underlying performance tradeoffs  and the associated  research challenges for future work. It was shown that  great opportunities  exist for the application of O-NOMA in the context of VLC, despite  certain restrictions imposed by the  nature  of VLC channel.  We believe that, with a  considerate   design of O-NOMA, VLC systems can  contribute towards meeting  the capacity  demands expected in future 5G  networks and beyond.

\bibliographystyle{IEEEtran}

\bibliography{magRef}

\begin{thebibliography}{10}
\providecommand{\url}[1]{#1}
\csname url@samestyle\endcsname
\providecommand{\newblock}{\relax}
\providecommand{\bibinfo}[2]{#2}
\providecommand{\BIBentrySTDinterwordspacing}{\spaceskip=0pt\relax}
\providecommand{\BIBentryALTinterwordstretchfactor}{4}
\providecommand{\BIBentryALTinterwordspacing}{\spaceskip=\fontdimen2\font plus
\BIBentryALTinterwordstretchfactor\fontdimen3\font minus
  \fontdimen4\font\relax}
\providecommand{\BIBforeignlanguage}[2]{{%
\expandafter\ifx\csname l@#1\endcsname\relax
\typeout{** WARNING: IEEEtran.bst: No hyphenation pattern has been}%
\typeout{** loaded for the language `#1'. Using the pattern for}%
\typeout{** the default language instead.}%
\else
\language=\csname l@#1\endcsname
\fi
#2}}
\providecommand{\BIBdecl}{\relax}
\BIBdecl

\bibitem{NOMA_VLC}
H.~Marshoud, V.~M. Kapinas, G.~K. Karagiannidis, and S.~Muhaidat,
  ``Non-orthogonal multiple access for visible light communications,''
  \emph{{IEEE} Photon. Technol. Lett.}, vol.~28, no.~1, pp. 51--54, Jan. 2016.

\bibitem{SC}
T.~Cover, ``Broadcast channels,'' \emph{IEEE Trans. Inf. Theory.}, vol.~18,
  no.~1, pp. 2--14, Jan. 1972.

\bibitem{NOMA_VLC2}
R.~C. Kizilirmak, C.~R. Rowell, and M.~Uysal, ``Non-orthogonal multiple access
  {(NOMA)} for indoor visible light communications,'' in \emph{Proc. 4th Int.
  Workshop on Optical Wireless Communications {(IWOW)}}, Sep. 2015, pp.
  98--101.

\bibitem{NOMA_VLC1}
L.~Yin, X.~Wu, and H.~Haas, ``On the performance of non-orthogonal multiple
  access in visible light communication,'' in \emph{Proc. IEEE 26th Annual
  International Symposium on Personal, Indoor, and Mobile Radio Communications
  {(PIMRC)}}, Aug. 2015, pp. 1354--1359.

\bibitem{NOMA_VLC3}
Z.~Yang, W.~Xu, and Y.~Li, ``Fair non-orthogonal multiple access for visible
  light communication downlinks,'' \emph{{IEEE} Wireless Commun. Lett.},
  vol.~6, no.~1, pp. 66--69, Feb. 2017.

\bibitem{NOMA_VLC4}
X.~Zhang, Q.~Gao, C.~Gong, and Z.~Xu, ``User grouping and power allocation for
  {NOMA} visible light communication multi-cell networks,'' \emph{{IEEE}
  Commun. Lett}, vol.~PP, no.~99, pp. 1--1, 2016.

\bibitem{nokia}
``Ultra dense network {(UDN)} white paper,'' \emph{Nokia}.

\bibitem{NOMA_VLC5}
X.~Guan, Y.~Hong, Q.~Yang, and C.~C.~K. Chan, ``Phase pre-distortion for
  non-orthogonal multiple access in visible light communications,'' in
  \emph{Proc. Optical Fiber Communications Conference and Exhibition {(OFC)}},
  Mar. 2016, pp. 1--3.

\bibitem{NOMA_haas}
L.~Yin, W.~O. Popoola, X.~Wu, and H.~Haas, ``Performance evaluation of
  non-orthogonal multiple access in visible light communication,'' \emph{{IEEE}
  Trans. Commun.}, vol.~64, no.~12, pp. 5162--5175, Dec. 2016.

\bibitem{NOMA_HANAA}
H.~{Marshoud}, P.~C. {Sofotasios}, S.~{Muhaidat}, and G.~K. {Karagiannidis},
  ``{On the Performance of Visible Light Communications Systems with
  Non-Orthogonal Multiple Access},'' \emph{ArXiv e-prints}, Dec. 2016.

\bibitem{MIMO_VLC_Haas}
T.~Fath and H.~Haas, ``Performance comparison of {MIMO} techniques for optical
  wireless communications in indoor environments,'' \emph{IEEE Trans. Commun.},
  vol.~61, no.~2, pp. 733--742, Feb. 2013.

\bibitem{marshoud}
H.~Marshoud, D.~Dawoud, V.~Kapinas, G.~K.~Karagiannidis, S.~Muhaidat, and
  B.~Sharif, ``{MU-MIMO} precoding for {VLC} with imperfect {CSI},'' in
  \emph{Proc. 4th International Workshop on Optical Wireless Communications
  {(IWOW)}}, Sep. 2015, pp. 93--97.

\bibitem{nonlinear}
K.~Ying, Z.~Yu, R.~J. Baxley, H.~Qian, G.~K. Chang, and G.~T. Zhou, ``Nonlinear
  distortion mitigation in visible light communications,'' \emph{IEEE Wireless
  Commun.}, vol.~22, no.~2, pp. 36--45, Apr. 2015.

\bibitem{massive-mimo1}
K.~Xu, H.~Yu, and Y.~J. Zhu, ``Channel-adapted spatial modulation for massive
  {MIMO} visible light communications,'' \emph{IEEE Photon. Technol. Lett.},
  vol.~28, no.~23, pp. 2693--2696, Dec. 2016.

\end{thebibliography}
\balance
\end{document}